\documentclass[
 preprint,
 superscriptaddress,
 amsmath,amssymb,
 aps,
 prl,
]{revtex4-1}

\usepackage{graphicx}
\usepackage{dcolumn}
\usepackage{bm}

\begin{document}


\title{Enhanced Strong Interaction between Nanocavities and p-shell Excitons Beyond the Dipole Approximation}

\author{Chenjiang Qian}
\author{Xin Xie}
\author{Jingnan Yang}
\author{Kai Peng}
\author{Shiyao Wu}
\author{Feilong Song}
\author{Sibai Sun}
\author{Jianchen Dang}
\author{Yang Yu}
\affiliation{Beijing National Laboratory for Condensed Matter Physics, Institute of Physics, Chinese Academy of Sciences, Beijing 100190, China}
\affiliation{CAS Center for Excellence in Topological Quantum Computation and School of Physical Sciences, University of Chinese Academy of Sciences, Beijing 100049, China}
\author{Matthew J. Steer}
\author{Iain G. Thayne}
\affiliation{School of Engineering, University of Glasgow, Glasgow G12 8LT, U.K.}
\author{Kuijuan Jin}
\author{Changzhi Gu}

\affiliation{Beijing National Laboratory for Condensed Matter Physics, Institute of Physics, Chinese Academy of Sciences, Beijing 100190, China}
\affiliation{CAS Center for Excellence in Topological Quantum Computation and School of Physical Sciences, University of Chinese Academy of Sciences, Beijing 100049, China}

\author{Xiulai Xu}

\email{xlxu@iphy.ac.cn}
\affiliation{Beijing National Laboratory for Condensed Matter Physics, Institute of Physics, Chinese Academy of Sciences, Beijing 100190, China}
\affiliation{CAS Center for Excellence in Topological Quantum Computation and School of Physical Sciences, University of Chinese Academy of Sciences, Beijing 100049, China}
\affiliation{Songshan Lake Materials Laboratory, Dongguan, Guangdong 523808, China}

\date{\today}

\begin{abstract}

Large coupling strengths in exciton-photon interactions are important for quantum photonic network, while strong cavity-quantum-dot interactions have been focused on s-shell excitons with small coupling strengths. Here we demonstrate strong interactions between cavities and p-shell excitons with a great enhancement by the \textit{in situ} wave-function control. The p-shell excitons are demonstrated with much larger wave-function extents and nonlocal interactions beyond the dipole approximation. Then the interaction is tuned from the nonlocal to local regime by the wave-function shrinking, during which the enhancement is obtained. A large coupling strength of $210\ \mu\mathrm{eV}$ has been achieved, indicating the great potential of p-shell excitons for coherent information exchange. Furthermore, we propose a distributed delay model to quantitatively explain the coupling strength variation, revealing the intertwining of excitons and photons beyond the dipole approximation.

\begin{description}
\item[PACS numbers]
42.50.Pq,78.67.Pt,78.67.Hc
\end{description}
\end{abstract}
\pacs{42.50.Pq,78.67.Pt,78.67.Hc }
\maketitle


Strong interactions between single photons and excitons in nanocavities play a central role in the quantum photonic network \cite{Kimble2008,Ritter2012}. The control and enhancement of exciton-photon interaction is significant to improve the efficiency and fidelity of the coherent control in quantum information processing \cite{Warburton2013,Carter2013,Bose2014}, thus the large coupling strength is always pursued in cavity quantum electrodynamics (CQED). Additionally, the control of coupling strength also provides the base for the study of many other exciton-photon interactions such as exceptional points and topological polaritons \cite{1751-8121-45-44-444016,PhysRevLett.104.153601,PhysRevX.5.031001}.

As an ideal material for the solid-state quantum photonic network, quantum dots (QDs) embedded in photonic crystal cavities provide exciton-photon polariton states with long coherence time and chip-scale integrability. However, previous investigations are mainly focused on the s-shell (ground state) with the dipole approximation (DA) uncritically adopted, limiting the coupling strength \textit{g} to a small value with low controllability \cite{RevModPhys.87.347}. The enhancement and control of \textit{g} by tuning the cavity mode or moving the emitter, which is valid for some specific materials \cite{GG2009,Birowosuto2014,PhysRevLett.119.233901,Gao2018}, requires complex mechanical controls and is unrealistic for the solid-state cavity-dot system. By contrast, the wave-function control by an external magnetic field can control the exciton-photon interaction \textit{in situ} \cite{PhysRevB.51.5570,PhysRevB.57.9088,PhysRevLett.103.127401}, but only a small decrease of \textit{g} has been obtained on s-shell with the DA \cite{PhysRevLett.103.127401,doi:10.1063/1.3562344,PhysRevLett.104.047402}.

Here we demonstrate the significant nonlocal interaction beyond the DA in the p-shell(excited state of QDs)-cavity system, which has a wave-function extent much larger than s-shell. The \textit{in situ} wave-function control is applied to tune the interaction from nonlocal to local regime. During the phase transition, the cavity-dot coupling strength is greatly enhanced with a largest value of $210\ \mu\mathrm{eV}$ achieved so far. The enhancement is quantitatively explained by a new phenomenological distributed delay model, which extends the local interaction in former monotonic decrease model \cite{PhysRevLett.103.127401} to the nonlocal interaction as a nontrivial intertwining of exciton and photon beyond the DA. Therefore, our work opens up a new area of excited states in QD based CQED with great significance to the solid-state quantum photonic network.

\begin{figure}
\includegraphics[scale=0.5]{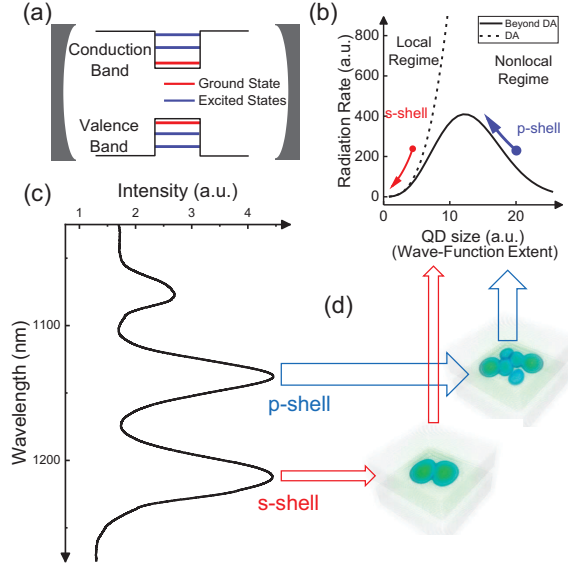}
\caption{\label{f1} (a) CQED system with a quantum emitter containing multiple excited states. (b) Radiation rate for QD with different wave-function extent under DA (dashed line) and beyond DA (solid line). Arrows shows the variation of coupling strength as the wave-function shrinking in magnetic field for s-shell (red) and p-shell (blue). (c) PL spectrum of ensemble QDs. The Gaussian peak at 1200 nm originates from s-shell and the peak at 1130 nm originates from p-shell. (d) Calculated wave-functions of hole states for s-shell and p-shell.}
\end{figure}

For the exciton-photon interaction between a quantum emitter with transition energy $\omega_{x}=\omega_{f}-\omega_{i}$ from the initial state $\left| i \right\rangle$ to the final state $\left| f \right\rangle$, and a quantized radiation field with cavity mode wave-function $\bm{\alpha}(\bm{r})$ (Fig.~\ref{f1}(a)), the perturbation theory gives the coupling strength \textit{g} proportional to $\left|\left\langle f\right| \bm{\alpha}(\bm{r})\cdot \bm{p} \left| i \right\rangle\right|$, where $\bm{p}$ is the momentum operator. As $\bm{\alpha}(\bm{r})$ is untunable for a solid-state nanocavity, the wave-function control on $\left| i \right\rangle$ and $\left| f \right\rangle$ is the only approach to the enhancement and control of \textit{g}. For quantum emitters, the wave-function can be modified by an external magnetic field. The magnetic field adds an additional lateral confinement with the magnetic length in the plane vertical to the field \cite{PhysRevB.51.5570,PhysRevB.57.9088,PhysRevLett.103.127401}. As the magnetic field increases, the additional confinement will narrow down and shrink the wave-function.

For excitons with wave-function extent much smaller than the photon wavelength, $\bm{\alpha}(\bm{r})$ could be considered as a constant and taken outside the integral. Then the interaction $\bm{\alpha}\cdot\left\langle f\right|\bm{p}\left|i\right\rangle\propto\bm{\alpha}\cdot\bm{d}$ is determined by the electric dipole moment $\bm{d}=\left\langle f\right| e\bm{r} \left| i \right\rangle$, known as the DA. The dipole moment $\bm{d}$ is related to the wave-function extent. Therefore, the interaction will decay with the wave-function shrinking, which has been demonstrated previously for the s-shell excitons \cite{PhysRevLett.103.127401,doi:10.1063/1.3562344}. However for excitons with large wave-function extent, $\bm{\alpha}(\bm{r})$ cannot be considered as a constant, thus exciton and photon cannot be separated and the nonlocal interaction beyond the DA becomes significant. Fig.~\ref{f1}(b) shows calculated nonlocal radiation rate of QDs with different size at the same wavelength based on the rigorous theory rather than dipole approximation or quadrupole approximation \cite{PhysRevB.86.085304}. The dashed line is the result with DA while the solid line beyond DA, and large QD size is equivalent to large wave-function extent. Although specific details may differ for various kinds of quantum emitters, the radiation rate generally will not infinitely increase with wave-function extent like the dashed line with DA. Additionally, the nonlocal effect is more significant in the cavity field \cite{PhysRevB.86.085304,RevModPhys.87.347}. For the radiation in homogeneous materials, the mode function of monochromatic plane-wave has a uniform density $|\bm{\alpha}(\bm{r})|^2$ with only phase difference. While in inhomogeneous materials such as cavities, $|\bm{\alpha}(\bm{r})|$ is non-uniform. In the photonic crystal cavity, $|\bm{\alpha}(\bm{r})|$ is large in the cavity center and small away from the center. Thus for the example of a quantum emitter in the cavity center, too large a wave-function extent obviously leads to the small coupling strength as results in Fig.~\ref{f1}(b), due to the small average value of $|\bm{\alpha}(\bm{r})|$.

\begin{figure}
\includegraphics[scale=0.5]{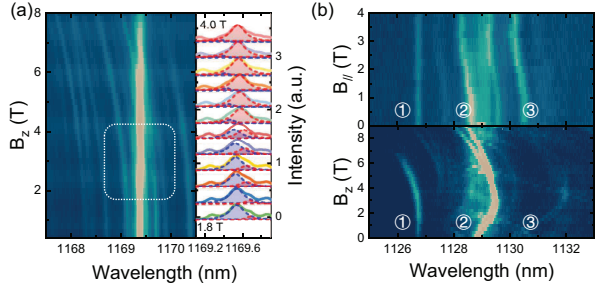}
\caption{\label{f2} (a) (left) PL map for s-shell transitions coupled to a high Q cavity mode in ${B_z}$. (right) An anti-crossing extracted from the dashed rectangular region in the left panel with coupling strength $g=45\ \mu\mathrm{eV}$, a typical coupling strength value for s-shell transitions, which is relatively small compared to p-shell transitions. (b) PL map for p-shell transitions in a low Q cavity mode in ${B_z}$, with diamagnetism reversal in ${B_z}$ (bottom) and in ${B_\parallel}$ (up) correspondingly. As marked in the figure, transition 1 has a normal positive diamagnetism in ${B_z}$ and a negligible diamagnetism in ${B_\parallel}$, while transition 2 and 3 have a diamagnetism reversal in ${B_z}$ and a relatively large diamagnetism in ${B_\parallel}$. The diamagnetic shift of p-shell is much larger than that of s-shell in (a), indicating a much larger wave-function extent.}
\end{figure}

Our sample contains a layer of self-assembled InAs QDs grown in the middle of a GaAs slab with a thickness of 170 nm. The detailed information of the sample and fabrication is in the Supplementary Information \cite{supplementary}. PL spectrum of ensemble QDs with large sizes indicates three main peaks for one ground state and two excited states (Fig.~\ref{f1}(c)). The s-shell and p-shell come from exciton recombination between the same electron state and two different hole states \cite{PhysRevB.59.5688}. The hole wave-function (Fig.~\ref{f1}(d)) of p-shell has a much larger extent than s-shell and even extends into the wetting layer \cite{PhysRevB.59.5688}, which can also be proved by the correlated diamagnetic shift \cite{PhysRevB.57.9088} shown in Fig.\ref{f2}. A few nonlocal interactions have been reported for the s-shell \cite{Andersen2010}, thus more significant nonlocal effect can be indicated from the larger wave-function extent of p-shell. Therefore, as the wave-function shrinks with the magnetic field, the p-shell-cavity interaction is continuously tuned from nonlocal regime to local regime. And the coupling strength variation can be predicted as the blue arrow in Fig.\ref{f1}(b), with the maximum value during the transition between the two regimes.


\begin{figure}
\includegraphics[scale=0.5]{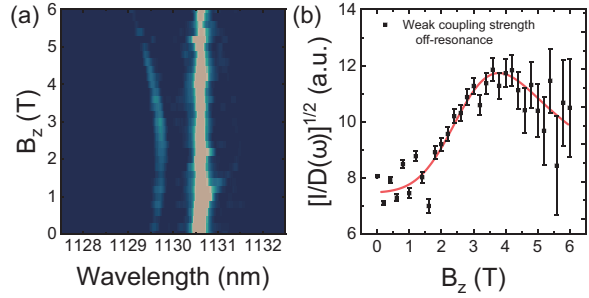}
\caption{\label{f3} (a) PL map of a p-shell transition with diamagnetism reversal in ${B_z}$. The transition is off-resonance to the cavity mode. (b) Coupling strength variation extracted by taking a square root of $I/D(\omega)$, with the fitting results by the EMG function (red line).}
\end{figure}

When the vertical magnetic field ${B_z}$ is applied, the diamagnetic shift of QD transitions is proportional to $\langle l_{\parallel}^2 \rangle B^2$, indicating the diamagnetism is related to the in-plane wave-function extent $l_{\parallel}$ \cite{PhysRevB.57.9088,PhysRevB.66.193303,PhysRevLett.103.127401}. Some p-shell transitions (bottom panel in Fig.~\ref{f2}(b)) have a diamagnetism reversal, negative below 3.5 T and positive above 3.5 T, different from other normal transitions. The reversal is difficult to be explained by the Fock-Darwin model with an invariable $l_{\parallel}$, which gives an abnormally big effective mass from the fitting result (See in the Supplementary Information \cite{supplementary}). In contrast, the reversal was explained with the shrinking of large wave-function extent as demonstrated previously \cite{PhysRevB.81.113307,Cao2015,Cao2016}. The wave-function extent of final state can be larger than the initial state due to the decrease of Coulomb attraction, resulting in the redshift when the wave-function of final state extends into the wetting layer. While with ${B_z}>3.5\ \mathrm{T}$, the emission peak is blueshifted as normal with further wave-function shrinking. The diamagnetism with a horizontal magnetic field ${B_\parallel}$ of these transitions is also larger than the normal transitions as well (upper panel in Fig.~\ref{f2}(b)), indicating large wave-function extent along the growth direction.

Due to the significant shrinking of the wave-function, the coupling strength \textit{g} of the p-shell-cavity system also varies with the magnetic field. In the weak coupling regime, Purcell enhanced spontaneous emission intensity of a quantum emitter in cavity radiation field with cavity mode $\omega_c$ and decay rate $\gamma_c=\omega_c/Q$ can be expressed by \cite{PhysRevB.60.13276,PhysRevLett.95.013904}
\begin{eqnarray}
\gamma_{SE}\propto{\left|\left\langle f\right| \bm{\alpha}(\bm{r})\cdot \bm{p} \left| i \right\rangle\right|^2}D_c(\omega_x)\nonumber
\end{eqnarray}
where $\left|\left\langle f\right| \bm{\alpha}(\bm{r})\cdot \bm{p} \left| i \right\rangle\right|$ is the coupling strength term, and $\pi D_c(\omega_x)=( \gamma_c/2 )/[{( \omega_x-\omega_c )^2+( \gamma_c/2 )^2}]$ is mode density term determined by the detuning. Fig.~\ref{f3}(a) shows PL map of an enhanced p-shell transition with diamagnetism reversal in ${B_z}$. The transition is around $1\ \mathrm{nm}$ off-resonance away from the cavity mode. The intensity of each peak was divided by the mode density $D_c(\omega_x)$ to focus on the coupling strength term (Fig.~\ref{f3}(b)). The coupling strength first increases with ${B_z}<3.5\ \mathrm{T}$ and then decreases with ${B_z}>3.5\ \mathrm{T}$. In contrast, only decrease of coupling strength can be predicted and observed if the DA is applied \cite{PhysRevLett.103.127401,doi:10.1063/1.3562344}. The increase of coupling strength directly proves the exciton-photon interaction beyond the DA, corresponding well with the nonlocal intearaction model (Solid line in Fig.~\ref{f1}(b)).

\begin{figure*}
\includegraphics[scale=0.7]{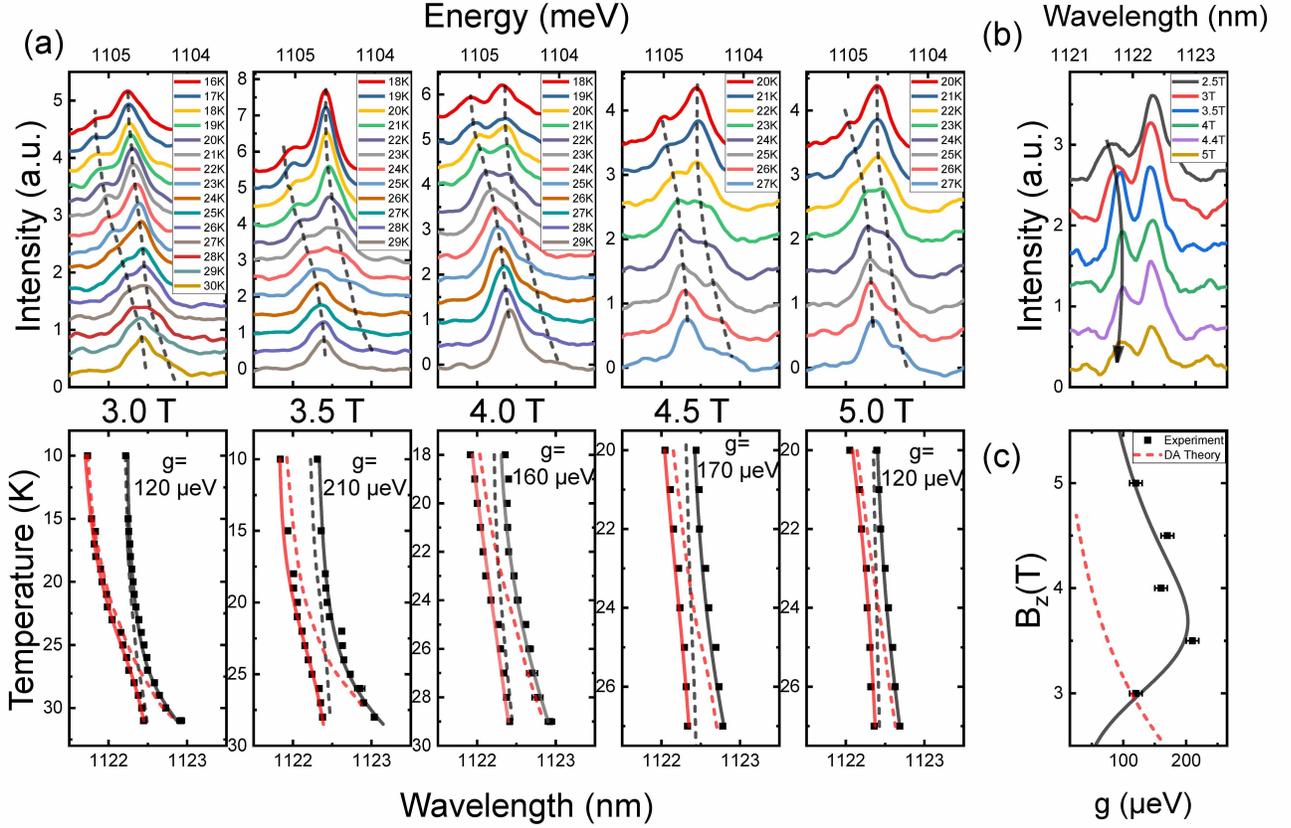}
\caption{\label{f4} (a) (up) Temperature-dependent PL spectra with anti-crossing refer to strong coupling between a p-shell transition and high Q cavity with a vertical magnetic field $3\ \mathrm{T}$, $3.5\ \mathrm{T}$, $4\ \mathrm{T}$, $4.5\ \mathrm{T}$ and $5\ \mathrm{T}$ as marked in the figure. (bottom) Fitted peak wavelength (black dot), bare cavity and peak wavelength (dashed lines) and fitting result by the strong coupling model (solid lines) corresponding to upper panels. The x axis is same for each panel, as the energy values shown upper and wavelength values shown bottom. (b) PL spectra of the p-shell transition in vertical magnetic field at $4.2\ \mathrm{K}$. (c) Coupling strength variation extracted from Rabi splittings, in good agreement with the EMG function (black solid line) refer to Fig.\ref{f3}(b), in contrast to the theoretical model with DA (red dashed line).}
\end{figure*}

\begin{figure}
\includegraphics[scale=0.8]{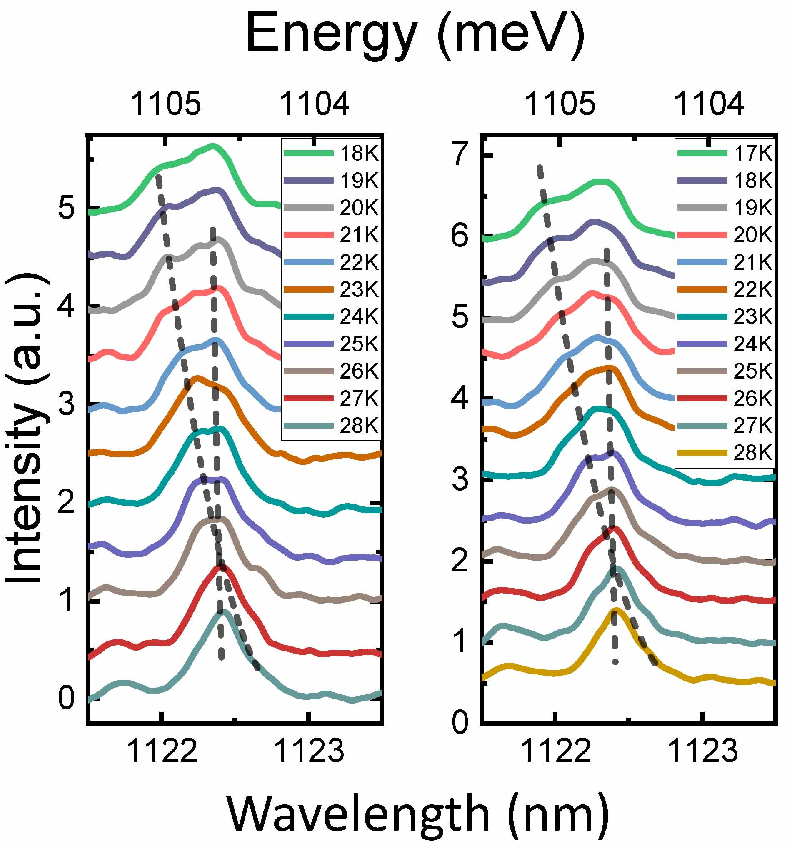}
\caption{\label{f5} Temperature-dependent PL spectra with a vector magnetic field $({B_\parallel},{B_z})$ of (left two panels) $(0.5\ \mathrm{T},3.5\ \mathrm{T})$ and (right panels) $(1.0\ \mathrm{T},3.5\ \mathrm{T})$. The interaction rapidly decays to the weak coupling regime with an additional ${B_\parallel}$.}
\end{figure}

In the strong coupling regime, the Rabi splitting on resonance is \cite{PhysRevB.60.13276,Hennessy2007}
\begin{eqnarray}
\Delta E=2\sqrt{g^{2}-\left(\frac{\gamma_x-\gamma_c}{4}\right)^2} \nonumber,
\end{eqnarray}
from which the coupling strength \textit{g} can be directly extracted. $\gamma_x$ ($\gamma_c$) is the decay rate of exciton (cavity). Fig.~\ref{f4} shows the PL spectra of a strongly coupled p-shell-cavity system. A p-shell transition nearby the cavity mode was observed with the similar reversal of diamagnetic shift and PL intensity (Fig.~\ref{f4}(b)), with the reversal points both around ${B_z}=3.5\ \mathrm{T}$. Series of temperature tuning were applied to tune the transition and cavity to resonance, with ${B_z}$ from 3 T to 5 T (upper panels in Fig.~\ref{f4}(a)). Then \textit{g} values were extracted from the well fitted results (bottom panels in Fig.~\ref{f4}(a)). The variation of \textit{g} (Fig.~\ref{f4}(c)) is in good agreement with results in the weak coupling regime (dark solid line) as expected. The maximum \textit{g} at ${B_z}=3.5\ \mathrm{T}$ is $210\ \mathrm{\mu eV}$ (Rabi splitting of $420\ \mathrm{\mu eV}$), much larger than the value achieved in s-shell-cavity system with analogous QDs \cite{PhysRevLett.120.213901}, and is also the largest value achieved in cavity-dot system so far \cite{doi:10.1063/1.5016615}. Additionally, the maximum \textit{g} rapidly decays to an unobservable value with a small additional ${B_\parallel}=0.5\ \mathrm{T}$ (Fig.~\ref{f5}), indicating a high controllability related to the large wave-function extent along the growth direction. Normally, a slower decay rate of coupling strength in ${B_\parallel}$ was observed for excitons with smaller wave-function extent (see in the Supplementary Information \cite{supplementary}).

The detailed calculation of wave-function in the magnetic field is non-trivial. Nonetheless, the coupling strength variation can be well explained by the wave-function shrinking. The former monotonic decay of coupling strength $f_{decay}(B)$ was explained with the decrease of dipole moment as wave-function shrinking with the DA. For the p-shell with large wave-function extent, we extend the former monotonic decay model to a decay model with distributed delay beyond the DA. The coupling strength is $\left|\left\langle f\right| \bm{\alpha}(\bm{r})\cdot \bm{p} \left| i \right\rangle\right|$, an integration of the coupling term at different positions. Meanwhile as ${B}$ increases, the additional confinement with magnetic length $\sqrt{{\hbar}/{eB}}$ narrows down, where ${\hbar}$ is the reduced Planck constant and $e$ is the elementary charge. This means the wave-function at $r'$ starts to shrink when $B'={\hbar}/{er'^2}$. This results in a delay of decay $f_{decay}(B-B')$ for wave-function at different $r'$ as ${B}$ increases, where $f_{decay}$ means the decay of wave-function in the magnetic field. Additionally, due to the nonlocal interaction, $\bm{\alpha}(\bm{r})$ is non-uniform. This means wave-function at different $\bm{r}$ has different contribution $f_{distribution}(\bm{r})$ to the coupling strength. Thus we can have a distributed delay model
\begin{eqnarray}
\hbar g(B)&=&\int f_{distribution}(r') f_{decay}(B-B')dr' \nonumber\\
&=&\int g_{distribution}(B')f_{decay}(B-B')dB' \nonumber.
\end{eqnarray}
$g_{distribution}(B)$ is the transform of $f_{distribution}(\bm{r})$ in the integration with $B={\hbar}/{er^2}$. The coupling strength variation in the experiment is well fitted by an exponentially modified Gaussian (EMG) function (solid lines in Fig.\ref{f3}(b) and Fig.\ref{f4}(c))
\begin{eqnarray}
f(x)&=&y_0+(f_1\otimes f_2)(x)\nonumber\\
f_1(x)&=&{A}e^{-\frac{x}{\tau}}\nonumber\\
f_2(x)&=&\frac{1}{\sqrt{2\pi}\sigma}e^{-\frac{(x-x_c)^2}{2\sigma^2}}\nonumber
\end{eqnarray}
where $f_1$ is an exponential decay and $f_2$ is a normal distribution. $f_1\otimes f_2(x)=\int f_1(x-z)f_2(z)dz$ is the convolution of two functions. The EMG function indicates an integration of exponential decay with distributed delay, and the delay has a normal distribution of weight, corresponding well with the distributed delay model. In contrast, for s-shell transitions with the DA, $\bm{\alpha}(\bm{r})$ is constant thus wave-function at different position has the same contribution, resulting in the degeneration back to a monotonic decrease of \textit{g} as reported previously \cite{PhysRevLett.103.127401}. The fitting results by EMG function in Fig.~\ref{f3}(b) are $\tau=3.7\ \mathrm{T}$ as the exponential decay rate, $x_c=2.5\ \mathrm{T}$ with corresponding magnetic length of 16 nm for the position with average contribution to \textit{g} and $\sigma=0.77\ \mathrm{T}$ with corresponding magnetic length of 5 nm for the standard deviation of the distribution. These values are in good agreement with the QD size. Therefore, the theoretical model well explains the coupling strength variation of both the s-shell with the DA in previous works and the p-shell beyond the DA in our experiment, revealing the nature of the transition between nonlocal and local interaction regimes.


In summary, we experimentally demonstrated the significant nonlocal interaction beyond the DA in the strongly coupled p-shell-cavity system and achieved great enhancement of the coupling strength. The magnetodynamics of the exciton-photon interaction is well described by the new distributed delay model. Our work makes it possible to enhance and control the single-exciton-photon interaction in solid state, which is a significant step to the build of quantum photonic network. Additionally, as the \textit{in situ} wave-function control is valid for other quantum emitters as well, this work can also be extended from single-exciton-photon interaction to new multi-dipole materials thus benefits various light-matter interactions such as biosensors and solar cells \cite{doi:10.1002/adma.201305583,Alivisatos2003}.

\begin{acknowledgments}
This work was supported by the National Basic Research Program of China under Grant No. 2016YFA0200400; the National Natural Science Foundation of China under Grant No. 11721404, 51761145104, 61675228 and 61390503; the Strategic Priority Research Program of the Chinese Academy of Sciences under Grant No. XDB07030200, XDB28000000 and XDB07020200; the Key Research Program of Frontier Sciences of CAS under Grant No. QYZDJ-SSW-SLH042; the Instrument Developing Project of CAS under Grant No.YJKYYQ20180036 and CAS Interdisciplinary Innovation Team. Authors would like to thank Gas Sensing Solutions Ltd for using the MBE equipment.
\end{acknowledgments}

\end{document}